\documentclass{aa}
\usepackage{txfonts}
\usepackage[latin1]{inputenc}
\usepackage{graphicx}
\usepackage{natbib}
\bibpunct{(}{)}{;}{a}{}{,} 

\begin{document}

\title{Performance study of ground-based infrared Bracewell interferometers}
\subtitle{Application to the detection of exozodiacal dust disks with GENIE}

\author{O.~Absil\inst{1}\thanks{O.~A. acknowledges the financial support of the Belgian National
Fund for Scientific Research (FNRS).} \and R.~den~Hartog\inst{2} \and P.~Gondoin\inst{2} \and P.
Fabry\inst{2} \and R.~Wilhelm\inst{3} \and P.~Gitton\inst{3} \and F.~Puech\inst{3}}

\institute{Institut d'Astrophysique et de G\'eophysique, Universit\'e de Liège, 17 All\'ee du Six
Ao\^ut, B-4000 Sart-Tilman, Belgium \\ \email{absil@astro.ulg.ac.be} \and Science Payloads and
Advanced Concepts Office, ESA/ESTEC, postbus 299, NL-2200 AG Noordwijk, The Netherlands \and
European Southern Observatory, Karl-Schwarzschild-Str. 2, D-85748 Garching bei München, Germany}

\offprints{O. Absil}

\date{Received 25 May 2005 / Accepted 2 November 2005}

\abstract{Nulling interferometry, a powerful technique for high-resolution imaging of the close
neighbourhood of bright astrophysical objets, is currently considered for future space missions
such as Darwin or the Terrestrial Planet Finder Interferometer (TPF-I), both aiming at Earth-like
planet detection and characterization. Ground-based nulling interferometers are being studied for
both technology demonstration and scientific preparation of the Darwin/TPF-I missions through a
systematic survey of circumstellar dust disks around nearby stars. In this paper, we investigate
the influence of atmospheric turbulence on the performance of ground-based nulling instruments, and
deduce the major design guidelines for such instruments. End-to-end numerical simulations allow us
to estimate the performance of the main subsystems and thereby the actual sensitivity of the nuller
to faint exozodiacal disks. Particular attention is also given to the important question of stellar
leakage calibration. This study is illustrated in the context of GENIE, the Ground-based European
Nulling Interferometer Experiment, to be installed at the VLTI and working in the L' band. We
estimate that this instrument will detect exozodiacal clouds as faint as about 50 times the Solar
zodiacal cloud, thereby placing strong constraints on the acceptable targets for Darwin/TPF-I.
 }

\titlerunning{Performance study of ground-based infrared Bracewell interferometers}
\authorrunning{Absil et~al.}

\maketitle

\section{Introduction}

The existence of extraterrestrial life in the Universe is a long-standing question of humankind.
Future space missions such as Darwin~\citep{Fridlund00} or the Terrestrial Planet Finder
Interferometer~\citep{BWL99,LD05} are currently being studied respectively by ESA and NASA to
search for evidence of life on Earth-like exoplanets by means of nulling
interferometry~\citep{Bracewell78}. These ambitious missions are particularly innovative from the
technology point of view and demand several techniques to be first tested on ground. Both ESA and
NASA have therefore initiated studies for ground-based infrared nulling interferometers (or
``nullers''), to be installed respectively at the VLTI~\citep{Glindemann04} or Keck
Interferometer~\citep{CW04}. A cryogenic nulling beam-combiner is also under development for the
LBT in the context of the NASA Origins Program~\citep{HH04}.

\begin{figure}[!h]
\centering
\includegraphics[width=7cm]{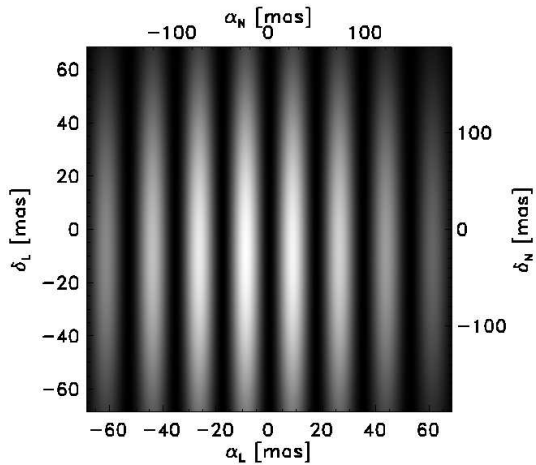}
\caption{Monochromatic transmission map for a 47~m Bracewell interferometer formed of two 8-m
telescopes. The interferometric field-of-view, limited by the use of single-mode waveguides (see
Sect.~\ref{sub:wfe}), is given by $\Omega = \lambda^2/S$ with $S$ the telescope surface. Its
diameter is respectively of 137~mas and 378~mas at the centers of the L' and N bands. The
transmission map acts as a ``photon sieve'': it shows the parts of the field that are transmitted
(white stripes) and those that are blocked (dark stripes, including the central dark fringe) by the
interference process. The whole transmitted flux is integrated on a single pixel.} \label{fig:tmap}
\end{figure}

The principle of nulling interferometry, first proposed by~\citet{Bracewell78} and generalized by
several authors~\citep{Angel89,Leger96,Angel97,Karlsson00,Absil02}, is to combine the light
collected by two or more telescopes in a co-axial mode, adjusting their respective phases in order
to produce a totally destructive interference on the optical axis. The interferometer is
characterized by its transmission map $T_{\lambda}(\theta,\phi)$, displayed in Fig.~\ref{fig:tmap}
for a two-telescope interferometer, in one arm of which a phase shift of $\pi$ radians has been
introduced ({\em Bracewell interferometer}). Its expression results from the combination of a
single telescope diffraction pattern with the fringe pattern produced by the interference between
the beams, i.e., for a two-telescope interferometer:
\begin{equation}
T_{\lambda}(\theta,\phi) = \left(\frac{2 J_1(\pi\theta D/\lambda)}{\pi\theta D/\lambda} \right)^2
\sin^2 \left(\pi \frac{B\theta}{\lambda} \cos\phi \right) \: ,
\end{equation}
where $\theta$ and $\phi$ are respectively the radial and polar angular coordinates with respect to
the optical axis, $B$ the interferometer baseline, $D$ the telescope diameter and $\lambda$ the
wavelength. In the following study, we assume that recombination and detection are both done in the
pupil plane, so that no image is formed: the flux is integrated on the field-of-view and detected
on a single pixel. Performing the detection in an image plane would not significantly change the
conclusions of this study.

Besides pure technology demonstration, ground-based nullers have an important scientific goal in
the context of the Darwin/TPF-I missions, through the detection of faint circumstellar dust disks
(``exozodiacal dust disks'') around nearby main sequence stars. Such dust clouds may present a
severe limitation to Earth-like planet detection with Darwin/TPF-I if they are more than 20~times
as dense as the Solar zodiacal cloud~\citep{Absil01}. A systematic survey of nearby main-sequence
stars is thus mandatory to select convenient targets for these future space missions. Photometric
surveys with infrared space telescopes such as IRAS, ISO and Spitzer have detected cold dust around
nearby stars at a level of $\sim 100$~times our Kuiper Belt, but did not allow the characterization
of warm dust emission at a better level than about 500~times the Solar zodiacal
cloud~\citep{Beichman05}. Attempts to spatially resolve faint exozodiacal clouds with single dish
telescopes in the mid-infrared~\citep{Kuchner98} and near infrared~\citep{Kuchner00} have not
yielded better detection limits. Even the MIRI instrument onboard the future James Webb Space
Telescope, equipped with coronagraphic devices at 10.6~$\mu$m and other longer wavelengths, does
not have a good enough angular resolution ($\sim 330$~mas) to detect warm dust within the habitable
zone of Darwin/TPF candidate targets.

Nulling interferometry, combining high dynamic range and high angular resolution, is particularly
appropriate to carry out this systematic survey of the Darwin/TPF-I candidate targets. The Darwin
target catalogue includes F, G, K and M-type main sequence stars up to 25~pc~\citep{Stankov05},
with median distances to these four stellar types of respectively 20, 20, 19 and 15~pc. Ideally,
the survey should detect exozodiacal clouds about 20~times as dense as the Solar zodiacal cloud
(``20-zodi'' clouds hereafter). Based on the model of \citet{Kelsall98}, such clouds are typically
$10^4$~times fainter than the star in the L' band (centered around 3.8~$\mu$m) or $10^3$~times
fainter in the N band (centered around 10.5~$\mu$m). The M band will not be considered in this
study, because is too much affected by water vapor to be useful for nulling interferometry
\citep{Young94}.

Besides this survey aspect, ground-based nullers have various interesting scientific applications.
Any target requiring both high dynamic range and high angular resolution will benefit from their
capabilities, including the detection and characterization of debris disks around Vega-type stars,
proto-planetary disks around Young Stellar Objects, high-contrast binaries, bounded brown dwarfs
and hot extrasolar giant planets. Extragalactic studies may also benefit from ground-based nulling
through the characterization of dust tori around nearby AGNs. Ground-based nullers are extremely
powerful at angular diameter measurements as well: thanks to their very high stability, they could,
for instance, detect pulsations in unresolved stars that still produce a non negligible stellar
leakage in the nulled data.

In this paper, we estimate the performance of ground-based infrared nulling interferometers for the
detection of exozodiacal dust around typical Darwin/TPF-I targets. The structure of the paper
follows the path of the light, beginning with the astrophysical sources and ending up after
detection with the data analysis procedures. In section~\ref{sec:nulling}, we present the different
contributors to the detected signals in ground-based nulling interferometry. Section~\ref{sec:turb}
then focuses on a particularly important contributor (instrumental leakage), which strongly depends
on the turbulent processes taking place in the Earth's atmosphere. We will show that this
stochastic contributor needs to be compensated in real time at the hardware level in order to reach
the required sensitivity. This has important consequences for the design of a ground-based nulling
interferometer, as discussed in section~\ref{sec:design}, where real-time control loops are
addressed. After passing though the instrument and being recorded by the detector, the signal still
needs to be carefully analyzed in order to extract the useful exozodiacal contribution from the raw
data. Section~\ref{sec:calib} describes the associated post-processing techniques. In all these
sections, we will take a Sun-like G2V star at 20~pc as a typical target for illustrative purposes.
Finally, these concepts are applied in section~\ref{sec:genie}, where we derive the expected
sensitivity of a specific instrument to the dust emission for different types of targets in the
Darwin star catalogue.


\section{Signals and noises in ground-based nulling interferometry} \label{sec:nulling}

    \subsection{Exozodiacal dust disk} \label{sub:exozodi}

The Solar zodiacal cloud, a sparse disk of 10-100~$\mu$m diameter silicate grains, is the most
luminous component of the solar system after the Sun. Its optical depth is only $\sim$10$^{-7}$,
but its integrated emission at 10~$\mu$m is about 300~times larger than the flux of an Earth-sized
planet. Very little is known about dust disks around main sequence stars. We will thus assume that
exozodiacal clouds have the same density distribution as the Solar zodiacal cloud, as described by
\citet{Kelsall98}, but with a different scaling factor. In the following study, we will also assume
exozodiacal disks to be smooth (no clump or wake in the disk) and to be seen face-on. With such
assumptions, the azimuth of the baseline has no influence on the observations: only the baseline
length matters, which is convenient for illustrative purposes. Nevertheless, we should note the
influence of the disk shape on the transmitted signal.
\begin{itemize}
\item The {\em disk inclination} has a strong influence on the transmitted amount of exozodiacal
light. If the disk is seen edge-on and oriented perpendicularly to the fringe pattern (i.e., major
axis of the disk parallel to the baseline), it will produce almost the same transmitted signal as a
face-on disk. However, if it is oriented in the same direction as the fringe pattern, a large part
of the disk emission will be cancelled out by the central dark fringe. The azimuth of the baseline
thus has a large influence on the detection of inclined disks, which is an advantage: if one lets
the projected baseline evolve as an effect of the Earth diurnal rotation, the exozodiacal disk
signal will be modulated as the baseline orientation changes, making the identification of the disk
signal easier. The effect will be all the larger when the disk inclination is close to edge-on
\citep{Absil03b}.
\item {\em Inhomogeneities} in the disk could lead to unexpected modulation as the baseline
changes. In fact, the exozodiacal dust emission is expected to be smooth (less than 1\% random
variations) except for rings and wakes due to gravitational trapping by planets or bands due to
recent asteroid or comet collisions \citep{Dermott98}. The detection of such features would be a
direct signature of the presence of planets or planetesimals in the disk. This would however
require a good signal-to-noise ratio since these structures will be much fainter than the
integrated disk emission.
\end{itemize}

    \subsection{Geometric stellar leakage} \label{sub:geomleak}

Even when the star is perfectly centered on the optical axis, a part of the stellar light still
leaks through the transmission map due to the finite extent of the stellar disk, an effect known as
geometric stellar leakage. We define the {\em nulling ratio}~$N$ as the ratio between the
transmitted stellar flux and the total stellar flux collected by the two telescopes. Assuming that
the stellar angular radius~$\theta_{\ast}$ is small as compared to the fringe spacing~$\lambda/B$,
and computing the transmitted flux as a two-dimension integral of the transmission map on the
stellar disk, one gets the following expression for the nulling ratio:
\begin{equation}
N = 1/\rho = \frac{\pi^2}{4} \left( \frac{B\theta_{\ast}}{\lambda} \right)^2 \: ,
\label{eq:startrans}
\end{equation}
assuming a uniform brightness across the stellar disk. The inverse of this quantity is called the
{\em rejection rate}~$\rho$, which is the fundamental figure of merit for a nulling interferometer.
Typical values of the rejection rate for a Bracewell interferometer observing a Sun-like star at
20~pc are given in Table~\ref{tab:rejrate}.
\begin{table}[h]
\begin{center}
\begin{tabular}{cccc}
\hline \hline Baseline & 14.4m & 47m & 85m
\\ \hline L' band (3.8~$\mu$m) & 22197 & 2084 & 637
\\        N band (10.5~$\mu$m) & 169573 & 15897 & 4863
\\ \hline
\end{tabular}
\caption{Rejection rate for a Bracewell interferometer observing a Sun-like star at a distance of
20~pc in the L' and N bands, for typical interferometric baselines (14.4~m = LBT, 47~m = VLTI, 85~m
= Keck). The star is assumed to be at zenith.} \label{tab:rejrate}
\end{center}
\end{table}

The rejection rate is quite sensitive to limb darkening, because it is precisely the outer parts of
the stellar disk that mostly contribute to stellar leakage. Assuming a simple linear limb-darkening
law $B_{\lambda}(\mu) = B_{\lambda}(1) [1-u_{\lambda}(1-\mu)]$ for the stellar surface brightness
$B_{\lambda}$, with $\mu$ the cosine of the angle between the normal to the surface and the line of
sight, the nulling ratio would be the following:
\begin{equation}
N_{\rm LD} = 1/\rho_{\rm LD} = \frac{\pi^2}{4} \left( \frac{B\theta_{\ast}}{\lambda} \right)^2
\left( 1- \frac{7u_{\lambda}}{15} \right) \: , \label{eq:rejrate}
\end{equation}
showing a linear dependance with respect to the limb-darkening parameter~$u_{\lambda}$. For a
typical K-band limb darkening $u_{\rm K}=0.26$ applicable to a Sun-like star \citep{CD95}, the
rejection rate would increase by a substantial 14\% assuming the same physical diameter in the
uniform and limb-darkened cases. Other potential contributors to the actual amount of stellar
leakage are the asymmetry of the stellar disk, through stellar oblateness and/or spots. However,
these are only second order effects for late-type main-sequence stars, which are generally close to
spherical symmetry and are expected to have only small and scarce spots. These effects will be
neglected in the discussion.

Due to the limited rejection rate, geometric leakage will often exceed the flux from the
exozodiacal disk at the destructive output of the nuller, especially in the L' band. Fortunately,
geometric leakage is mostly deterministic, allowing the prediction of its contribution. This
calibration procedure is discussed in Sect.~\ref{sub:geomcal}.

    \subsection{Instrumental stellar leakage} \label{sub:instleak}

The expression~(\ref{eq:rejrate}) of the rejection rate is valid only for a perfect Bracewell
interferometer. In practice, the rejection rate is degraded by atmospheric turbulence and various
instrumental effects causing imperfect co-phasing of the light beams, intensity mismatches and
polarization errors~\citep{Ollivier99}. This contribution, called {\em instrumental leakage}, adds
to the geometric leakage at the destructive output of the interferometer. It is dominated by
non-linear, second order error terms and does not depend on the stellar diameter to the first
order~\citep{Lay04}.

Instrumental leakage has two effects on the performance of the nuller: first, it introduces a {\em
bias}, the mean contribution of instrumental leakage, and second, it introduces an {\em additional
stochastic noise} through its fluctuations. While the second contribution can be reduced by
increasing the observation time (see Sect.~\ref{sub:perf}), the first one does not improve with
time and limits the global performance of the nuller. Unlike geometric leakage, an analytical
expression of instrumental leakage cannot be obtained: it depends on the particular shape and
amplitude of the power spectral densities of various atmospheric effects and instrumental errors
such as piston, dispersion, wavefront errors, polarization errors, etc~\citep{Lay04}. Its reduction
can thus only be done through real-time control of these errors. In practice, instrumental leakage
can be at least partially calibrated by observing ``calibrator stars'' with well-known diameters, a
common method in stellar interferometry. This calibration, discussed in Sect.~\ref{sub:instcal},
will help remove the bias associated with the mean instrumental leakage and thereby relax the
requirements on the performance of the real-time control loops.

    \subsection{The sky background}

\begin{figure}[!h]
\centering \resizebox{\hsize}{!}{\includegraphics{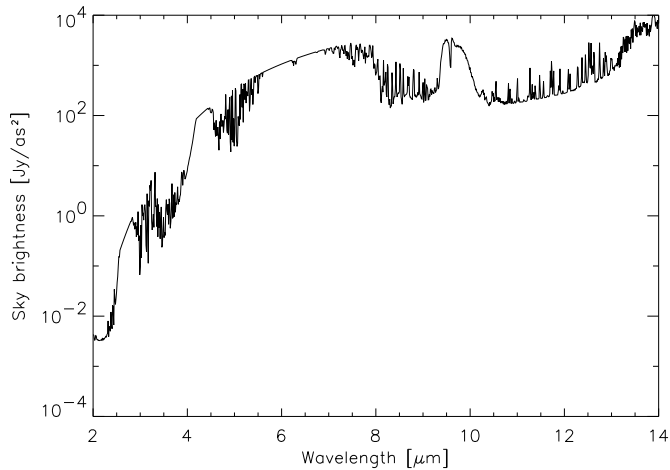}} \caption{Infrared sky brightness
for typical Paranal atmospheric conditions.} \label{fig:sky}
\end{figure}

The thermal emission of the sky, which peaks in the mid-infrared, is obtained in good approximation
by multiplying the Planck blackbody function by the wavelength-dependent sky emissivity. The
infrared sky brightness is plotted in Fig.~\ref{fig:sky} for an atmosphere at a mean temperature of
284~K, which is typical for Cerro Paranal. The mean sky brightness in the relevant atmospheric
wavebands is given in Table~\ref{tab:bckg}.

The fluctuation of the infrared background radiation produced by the atmosphere (or {\em sky
noise}) has not been well modelled to date. Sky noise measurements have been carried out in the N
band at Paranal with the VLTI mid-infrared instrument MIDI~\citep{Absil04}, while L'-band
measurements have been carried out at Siding Springs, Australia~\citep{Allen81}, showing that, at
low frequencies, sky noise dominates over the shot noise associated with the mean background
emission. Typical logarithmic slopes ranging from $-1$ to $-2$ have been recorded for the sky noise
power spectral density (PSD).

    \subsection{The instrumental background}

In order to conduct the light beams from the telescopes to the instrument, a large number of relay
optics are used. All of them emit at infrared wavelengths. The wavelength-dependent emissivity of
the optical train is theoretically equal to the complementary of its transmission. In order to
evaluate this contribution, we have to make some assumptions about the interferometric facility and
nulling instrument. We will take the case of the VLTI, which has a transmission ranging between
30\% and 40\% in the near- and mid-infrared\footnote{Due to the additional emission of dust
particles located on the optics, the measured emissivity of the VLTI optical train is in fact
closer to 100\% than the expected 60-70\%.}, and we will assume that the nulling instrument is at
ambient temperature, except for the detection unit (spectrograph and detector). The instrumental
background is given in Table~\ref{tab:bckg}, separated into ``VLTI brightness'' for the
interferometric infrastructure and ``GENIE brightness'' for the nulling instrument, assuming an
emissivity of 30\% for the latter. A full cryogenic instrument would reduce the background
emission, and thus improve the global sensitivity. However, even at ambient temperature, the full
potential of the instrument can still be achieved by increasing the integration time, thereby
reducing the contribution of the shot noise associated with the background emission. Fluctuations
of the instrumental background are also expected to happen (e.g.\ due to beam wandering, mirror
vibrations, gain fluctuations, etc.), but at a much lower level than sky noise. They will be
neglected in the following discussion.

    \subsection{Detection criterion} \label{sub:criterion}

In addition to the classical sources of noise such as shot noise, read-out noise and background
noise, whose effect can be reduced by increasing the integration time, nulling interferometry is
faced with two types of biases that must be removed to ensure a secure detection of the exozodiacal
cloud. These biases are associated with the mean contribution of geometric and instrumental stellar
leakage. In order to detect circumstellar features as faint as $10^{-3}$ (N band) to $10^{-4}$ (L'
band) of the stellar flux with a signal-to-noise ratio of 5, {\em one should reduce the
contributions of both geometric and instrumental stellar leakage down to $10^{-4}$ (N band) or
$10^{-5}$ (L' band) of the initial stellar flux}, so that the total stellar leakage does not exceed
about one fifth of the exozodiacal disk signal. Real-time control systems at hardware level and
calibration techniques at software level will be used to reach this level of performance. The
requirements on phase and intensity control will be particularly tight for short operating
wavelengths. In the next section, the discussion of atmospheric turbulence is thus mainly
illustrated in the L' band where its effects are the largest.


\section{Influence of atmospheric turbulence} \label{sec:turb}

In order to estimate the performance of a nulling interferometer in the presence of atmospheric
turbulence, we have developed a software simulator called {\em GENIEsim}\footnote{GENIEsim, written
in IDL, is open source and available on demand. Send request to R.~den Hartog ({\tt
rdhartog@rssd.esa.int}).}~\citep{Absil03a}. This semi-analytical simulator computes in real-time
the transmission map of the interferometer, taking into account phase and intensity fluctuations
associated with atmospheric turbulence. It inherently takes into account the bilinear error terms
discussed by~\citet{Lay04} that are supposed to be the largest contributors to the instrumental
nulling. In the following paragraphs, we describe the various sources of phase, intensity and
polarization errors that have been implemented in the simulator and discuss their effect on
instrumental nulling.

    \subsection{The piston effect} \label{sub:piston}

\begin{figure}[!h]
\centering \resizebox{\hsize}{!}{\includegraphics{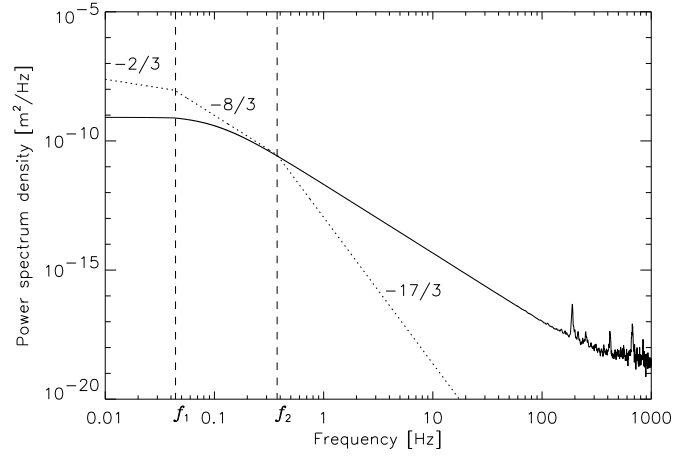}} \caption{Differential piston power
spectrum, under the following conditions: $B=47$~m, $D=8$~m, $v=11$~m/s and $r_0=10$~cm at 500~nm
(equivalent to 1'' seeing). The dotted line represents the theoretical Kolmogorov piston PSD,
showing the effects of pupil correlations (below $f_1 \simeq 0.2v/B$) and pupil averaging (above
$f_2 \simeq 0.3 v/D$), but without the effect of the outer scale nor the instrumental
contributions. The solid line is the representation of the piston PSD that we actually use in our
simulations as a conservative scenario. It includes the effect of pupil correlations and of the
outer scale of turbulence (flattening of the PSD for frequencies below $v/{\cal L}_0$), as well as
the contributions of bimorph and coupled piston (high-frequency fluctuations), but excludes pupil
averaging. A best-case scenario would correspond to the combination of the dotted line for $f> f_2$
and the solid line for $f<f_2$.} \label{fig:opdpsd}
\end{figure}

The 0th-order term of atmospheric turbulence, the spatially averaged phase perturbation on the
pupil, is known as the piston effect. Optical path difference (OPD) fluctuations are mainly due to
differential piston between the two apertures of the interferometer. Under the Taylor hypothesis of
frozen turbulence, where a frozen phase screen is blown at the wind speed above the
telescopes~\citep{Roddier81}, the classical description of atmospheric turbulence proposed
by~\citet{Kolmogorov41} states that the temporal PSD of phase fluctuations follows a $-8/3$
power-law. In practice, three physical phenomena induce a departure from the theoretical $-8/3$
logarithmic slope~\citep{Conan95}: correlations between the pupils and the finite outer scale of
turbulence ${\cal L}_0$ both reduce the fluctuations at low frequencies, while pupil averaging
reduces them at high frequencies (Fig.~\ref{fig:opdpsd}). The latter effect has in fact never been
observed, so that we keep the $-8/3$ slope at high frequencies as a conservative scenario. This
scenario also features additional instrumental contributions:
\begin{itemize}
\item Bimorph piston effect: this is the OPD induced by resonances of the adaptive optics
deformable mirror~\citep{VC01}. This contribution has been evaluated for the MACAO adaptive optics
at VLTI, and turns out to be dominant above 100~Hz despite the use of a ``piston-free'' algorithm
for the control of the deformable mirror.
\item Coupled piston: this is the additional OPD created by high-order wavefront errors when
injecting the beams into single-mode waveguides~\citep{RC01}.
\end{itemize}

The standard deviation of the OPD fluctuations, neglecting the effect of the outer scale, is given
by~\citep{Roddier81}:
\begin{equation}
\sigma_{\rm OPD} = 2.62\frac{\lambda}{2\pi} \left(\frac{B}{r_0}\right)^{5/6} \: .
\end{equation}
Since the Fried parameter $r_0$ is proportional to $\lambda^{6/5}$, $\sigma_{\rm OPD}$ does not
depend on wavelength. For a typical baseline of 47~m and 1'' seeing, the theoretical variance of
piston is about 35~$\mu$m, i.e., about 9 fringes in the L' band. This variance, which is also the
area under the dotted curve in Fig.~\ref{fig:opdpsd}, sets the level of the PSD.

The effect of phase errors on a Bracewell nulling interferometer is to shift the position of the
fringes in the transmission map: the dark fringe is not centered any more on the optical axis. For
a small phase error $\epsilon_{\phi}(\lambda)$ between the two beams, the shifting of the dark
fringe produces a non-null on-axis transmission $T_{\lambda}(0,0) = \epsilon_{\phi}(\lambda)^2/4$
\citep{Ollivier99}. In order to keep the instrumental rejection rate above the required $10^5$ in
the L' band, the phase error should be smaller than 0.006~radian (i.e., 4~nm in the L' band), while
atmospheric piston induces OPD errors of about 35~$\mu$m. A deep and stable null can thus only be
achieved by stabilizing the dark fringe on the optical axis, by means of a fringe tracker
(Sect.~\ref{sub:tracking}).

    \subsection{Longitudinal dispersion} \label{sub:disp}

\begin{figure}
\centering \resizebox{\hsize}{!}{\includegraphics{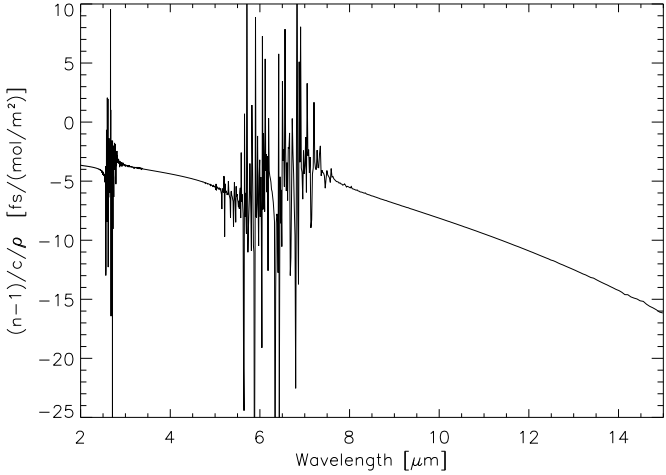}} \caption{Reduced refraction index
of WDA ($\hat{n}_{\rm WDA} = \frac{n_{\rm WDA}-1}{c\rho}$, with $c$ the speed of light and $\rho$
the molar density) expressed in femtosecond per mole/m$^2$, plotted at infrared wavelengths.}
\label{fig:wda}
\end{figure}

In addition to the achromatic piston effect produced by the fluctuations of the air refractive
index, another source of phase errors comes from the fluctuation of the water vapor column
densities above the two telescopes. A useful concept to describe the column density fluctuations of
water vapor is ``Water-vapor Displacing Air'' (WDA), introduced by~\citet{Meisner02}. The
fluctuation of the WDA column density induces a chromatic OPD difference between the beams, because
its refraction index $n_{\rm WDA}$ strongly depends on wavelength in the infrared as illustrated in
Fig.~\ref{fig:wda}. This effect, referred to as {\em longitudinal dispersion}, has the same kind of
influence on the transmission map as the piston effect, except that the offset of the dark fringe
is now wavelength-dependent. This has two deleterious consequences on the performance of a nulling
instrument:
\begin{itemize}
\item {\em Inter-band dispersion} is the phase error measured at the center of the science
waveband assuming that piston is perfectly corrected at another wavelength by means of a fringe
tracker (this correction is generally done in the H and/or K band, e.g.\ at the Keck and VLT
interferometers),
\item {\em Intra-band dispersion} refers to the differential phase error within the science
waveband: if one manages to co-phase the beams at the central wavelength of the science waveband,
their phases will not be perfectly matched at the edges of the band.
\end{itemize}

Since longitudinal dispersion is produced by the same turbulent atmosphere as piston, it is
expected to follow the same theoretical PSD. This has been confirmed experimentally
by~\citet{Lay97}. An RMS value of 1.5~mole/m$^2$ for the column density fluctuation of water vapor
has been measured by~\citet{Meisner02} at Paranal on an integration time of 100~sec (this value
corresponds to 27~$\mu$m RMS of differential precipitable water vapor). This translates into
typical inter-band and intra-band OPD errors of 550~nm and 150~nm RMS respectively in the L' band
assuming that fringe tracking is performed in the H band. A control system must thus be devised to
stabilize the dark fringe on the optical axis in order to meet the requirements on the instrumental
rejection rate.

    \subsection{Wavefront errors} \label{sub:wfe}

Propagation through the turbulent Earth atmosphere degrades the shape of the wavefronts, producing
imperfect interference between the beams at recombination, and thus creates a halo of incoherent
stellar light at the detector. Modal filtering by means of single-mode fibers, as proposed
by~\citet{Mennesson02}, eliminates wavefront corrugations by projection on the fundamental mode of
the fiber, thereby ensuring perfect matching of the wavefronts over a broad bandpass. After
injection into a single-mode fiber, the shape of the initial wavefront only affects the amount of
energy coupled into the guide. {\em The modal filter thus converts phase defects into intensity
errors}, which are less severe for a nulling interferometer~\citep{Ollivier99}. It also induces a
small coupled piston due to the non-null mean of $z>1$ Zernike polynomials projected onto the
Gaussian fundamental mode of the fiber~\citep{RC01}. The use of single-mode fibers to filter the
beams at the focus of the telescopes strongly relaxes the constraints on pointing accuracy and low
order aberrations, and thus improve the capability to achieve deep rejection ratios. We will assume
that modal filtering, or at least spatial filtering using pinholes for the N band, is used in our
nulling interferometer.

\begin{figure}[!h]
\centering \resizebox{\hsize}{!}{\includegraphics{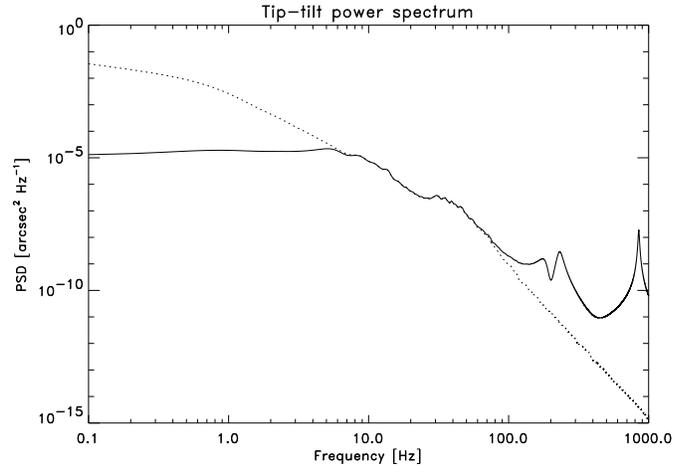}} \caption{PSD for tip-tilt before
(dotted line) and after (solid line) close-loop control with MACAO. The peaks at high frequencies
are induced by mechanical resonances in the deformable mirror.} \label{fig:psdtilt}
\end{figure}

In the following study, we will also assume that the telescopes are equipped with adaptive optics,
with performance similar to that of the MACAO systems on ESO's Very Large
Telescopes~\citep{Arsenault02}. The power spectral densities of individual Zernike modes are taken
from~\citet{Conan95}, with typical high-frequency slopes of $-17/3$ and various slopes at low
frequencies. These PSDs are corrected by the adaptive optics servo system for frequencies below
about 10~Hz, thereby making their low-frequency content negligible. The case of tip-tilt is
illustrated in Fig.~\ref{fig:psdtilt}, including the contribution of high order Zernike
modes~\citep{tenBrummelaar95}. The values for the RMS tip-tilt and Strehl are given in
Table~\ref{tab:macao}.

The coupling efficiency $\eta$ into a single-mode fiber is computed by taking into account two
contributions: the tip-tilt error which induces an offset of the stellar Airy pattern with respect
to the fiber core (assuming the on-sky tip-tilt to be perfectly translated into an on-fiber
jitter), and the Strehl ratio which represents the coherent part of the beam and can be taken as a
multiplicative factor in the computation of coupling efficiency~\citep{RC01}. The mean and RMS
coupling efficiency after correction by adaptive optics are given in Table~\ref{tab:macao}.

The effect of coupling efficiency fluctuations, which produce unequal intensities in the two arms
of the interferometer, is to induce a non-null transmission on the optical axis. Letting
$\epsilon_{I}$ be the relative intensity error between the two beams, the expression of the central
transmission is $T_{\lambda}(0,0)= \epsilon_I^2/16$~\citep{Ollivier99}. In order to keep the
instrumental rejection rate above the required $10^5$, relative intensity errors should be smaller
than 1\%, while the AO-corrected atmospheric turbulence induces intensity fluctuations as high as
8\% in the L' band. A servo loop for intensity matching is therefore necessary to obtain a deep and
stable null in the L' band. Such a device is not needed in the N band, where relative fluctuations
of 1\% are expected.

\begin{table}[!h]
\begin{center}
\begin{tabular}{cccc}
\hline \hline & Average & Std deviation
\\ \hline Tip-tilt [mas] & 0 & 14
\\ Strehl L' band & 0.80 & 0.050
\\ Strehl N band & 0.97 & 0.008
\\ \hline Coupling L' band & 0.58 & 0.044
\\ Coupling N band & 0.77 & 0.008
\\ \hline
\end{tabular}
\caption{RMS tip-tilt, mean and RMS Strehl ratio for an 8-m telescope after correction by MACAO, as
simulated by GENIEsim under typical atmospheric conditions at Cerro Paranal (1'' seeing, 11 m/s
wind speed). The mean and RMS coupling efficiency into a single-mode fiber is deduced (taking into
account the effect of central obscuration).} \label{tab:macao}
\end{center}
\end{table}

    \subsection{Scintillation}

Scintillation is the effect of rapid intensity fluctuations of a point-like source as a result of
the interference of light rays diffracted by turbulent cells. The condition for scintillation to
take place is that the Fresnel scale $r_{\rm F} = (\lambda \, h \, {\rm sec}z)^{1/2}$ is larger
than the Fried scale $r_0$, with $h$ the height of the turbulent layer and $z$ the zenith
distance~\citep{Quirrenbach99}. Even for high turbulent layers (20~km) and large zenith distances
($z=45$°), we obtain $r_{\rm F} \simeq 0.3$~m while $r_0 \simeq 1.1$~m in the L' band for 1''
seeing, showing that the Fried parameter is always larger than the Fresnel scale. Numerical
estimates obtained with GENIEsim confirm this statement: scintillation does not induce intensity
fluctuations larger than 0.2\% in the L' band. Since $r_0 \propto \lambda^{6/5}$ while $r_{\rm F}
\propto \lambda^{1/2}$, scintillation is also negligible for longer wavelengths, including the N
band.

    \subsection{Polarization errors}

The polarization errors encountered by a nulling interferometer are of three main types:
differential phase shift, differential attenuation and differential rotation can occur between the
two beams for each linear polarization component. These effects can be translated into phase and
intensity errors, but unlike phase and intensity errors due to turbulence, they are only due to the
instrument itself. Polarization errors are thus expected to be mainly static or slowly drifting,
even if instrumental vibrations might induce some fluctuations. A highly symmetric design of the
whole interferometer is required to reduce polarization issues as much as possible~\citep{SC01}. In
the following study, we will assume that the nulling instrument is designed so as to keep this
contribution negligible in the instrumental leakage budget. This may not be the case in practice,
because existing interferometers are not necessarily designed to meet the tight tolerances on
polarization associated with nulling interferometry. This might force the nuller to operate in
polarized light, by inserting polarizers at the beginning of the optical train\footnote{In
practice, a polarizer cannot be put in front of the primary or secondary mirror of a telescope, so
that the first few mirrors of the optical train could still contribute to polarization errors.} and
just before recombination. The instrumental throughput would then be reduced by 50\%, thereby
increasing the required integration times.


\section{Real-time control and instrumental design} \label{sec:design}

We have seen that atmospheric turbulence causes additional stellar photons to leak through the null
({\em instrumental leakage}), and that this bias cannot be calibrated analytically as for geometric
leakage. Moreover, the stochastic fluctuation of instrumental leakage is the source of an important
noise contribution, called {\em systematic noise}~\citep{Lay04} or {\em variability
noise}~\citep{CB05}. This section describes the three major control systems that are needed to
reduce instrumental leakage and variability noise down to an acceptable level by correcting the
effects of atmospheric turbulence in real time. As discussed in Sect.~\ref{sec:turb} and summarized
in Table~\ref{tab:nullperf}, an instrumental rejection rate of $10^5$ requires the residual phase
and intensity errors to be smaller than 4~nm and 1\% RMS respectively. If this level of performance
cannot be met, the use of calibrator stars will be further investigated to estimate and remove the
contribution of instrumental leakage.

    \subsection{Fringe tracking} \label{sub:tracking}

\begin{figure}[!h]
\centering
\includegraphics[width=7cm]{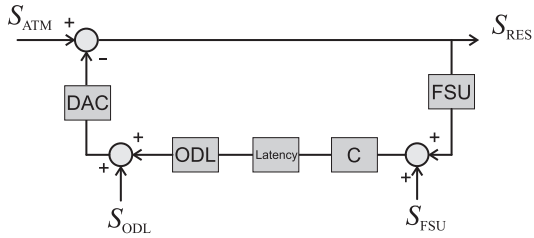}
\caption{Simplified block-diagram of a fringe tracking system: the OPD error between two beams,
characterized by its PSD $S_{\rm ATM}$, is sampled by a fringe sensing unit (FSU) feeding an OPD
controller (C), which computes the correction to be applied by the optical delay line (ODL). The
latency represents the effect of the finite CPU and electronics time required between fringe
sensing and correction, while the digital-to-analogue converter (DAC) represents the conversion
from a digital command to an analogue OPD correction by the delay line. The OPD power spectrum
after closed-loop control is denoted $S_{\rm RES}$, while the noise introduced by the FSU and the
ODL are respectively $S_{\rm FSU}$ and $S_{\rm ODL}$.} \label{fig:fsublock}
\end{figure}

The purpose of a fringe tracker is to actively compensate for random phase fluctuations between the
beams of the interferometer at a given wavelength. A fringe sensing unit (FSU) measures the phase
difference at a given sampling frequency and transmits this information to a controller which
computes the compensation to be applied to one of the beams by an optical delay line (ODL), as
illustrated in Fig.~\ref{fig:fsublock}. Both the FSU and the ODL introduce noise in the fringe
tracking control loop, while the controller is (almost) noiseless. In order to be efficient, fringe
tracking requires a high signal-to-noise ratio to be achieved on phase measurements on very short
time scales (typically between 0.1 and 1~msec). It is therefore usually performed in the
near-infrared ($\lambda<2.4$~$\mu$m) where stellar photons are numerous and the sky background
rather low.

\begin{figure}[!h]
\centering \resizebox{\hsize}{!}{\includegraphics{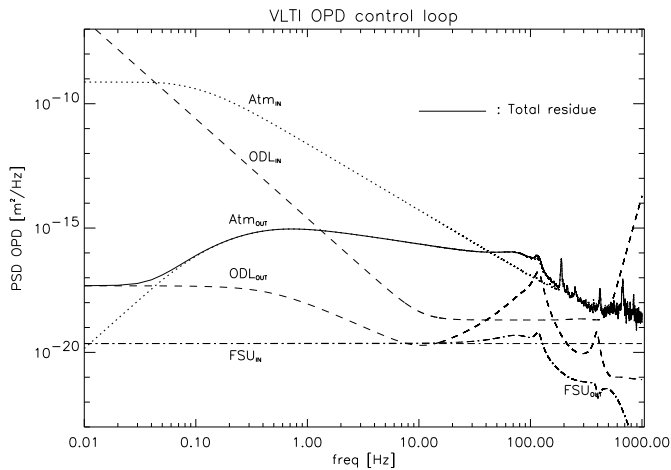}} \caption{Simulation of the VLTI OPD
control loop. The plot gives the PSDs at the input (before actuation: subscript ``IN'') and output
(after actuation: subscript ``OUT'') of the loop for the three noise sources in the VLTI fringe
tracking loop, i.e., atmospheric fluctuations (Atm, dotted curves), fringe sensing (FSU,
dashed-dotted curves) and optical delay line (ODL, dashed curves). The total closed-loop OPD jitter
(area under the solid line) is 133~nm RMS for a repetition frequency of 4~kHz.} \label{fig:psdfsu}
\end{figure}

Large interferometric facilities such as the VLTI or Keck Interferometer are (or will soon be)
equipped with fringe trackers. For instance, FINITO and PRIMA at VLTI will deliver a residual OPD
of about 150~nm RMS on bright stars, respectively in the H and K bands~\citep{Wilhelm03}. However,
this is not sufficient to meet our goal on OPD stability, implying that a second stage of OPD
control must be implemented inside the nulling instrument. In fact, the main limitation to the
accuracy of the VLTI fringe tracker comes from the ODL, which does not efficiently correct for OPD
fluctuations at frequencies higher than 50~Hz (see Fig.~\ref{fig:psdfsu}). The main requirement for
the additional fringe tracker is thus to be equipped with a low-noise ODL optimized for high
repetition frequencies (up to 20~kHz).

We have simulated the performance of a fringe tracker optimized to achieve small residual OPD on
bright stars. The fringe tracker features a fringe sensor working both in the H and K band and uses
a short-stroke fast piezo or voice-coil delay line to compensate the remaining OPD fluctuations.
The control loop parameters are computed through a simultaneous optimization of the repetition
frequency and of the controller parameters (a simple PID\footnote{PID stands for ``Proportional,
Integral and Differential'' and is a basic controller for closed-loop control.}). The repetition
frequency has been artificially limited to 20~kHz to comply with state-of-the-art back-end
electronics. The delay line is assumed to have a perfect response up to this frequency
(short-stroke piezo translators can achieve this kind of performance). The OPD control performance,
summarized in Table~\ref{tab:nullperf}, is computed in two cases:
\begin{itemize}
\item the {\em worst case} corresponds to a $-8/3$ slope for the turbulence PSD at high
frequencies (see Fig.~\ref{fig:opdpsd}) and includes the bimorph piston effect;
\item the {\em best case} takes into account the effect of pupil averaging on the PSD ($-17/3$
slope at high frequencies) and does not include the bimorph piston effect.
\end{itemize}
The predicted performance ranges between 6~nm and 17~nm RMS. These figures could be improved by
using a priori knowledge on the behaviour of atmospheric turbulence in the control process, e.g.\
through Kalman filtering. First estimates have shown an improvement of about 2~nm on the residual
OPD with respect to the above numbers (C.~Petit, personal communication). This performance,
summarized in Table~\ref{tab:nullperf}, is marginally compliant with the required 4~nm RMS.

    \subsection{Compensation of longitudinal dispersion}

There are two different sources of dispersion to be corrected: the slowly varying differential
phase due to unequal paths through wet air in the delay lines, and an additional rapid phase
fluctuation due to column density fluctuations of dispersive components in the atmosphere. The
first one is deterministic and easy to correct provided that the atmospheric conditions are
monitored in the delay lines, while the second one is the real problem and is mainly associated
with water vapour, as well as other molecular species such as ozone, carbon dioxide, etc, which are
the main contributors below 2.5~$\mu$m~\citep{Mathar04}.

In order to correct for both inter-band and intra-band dispersion (Sect.~\ref{sub:disp}), one
should measure the differential column density of water vapour experienced by the two beams in
their paths through the atmosphere. This quantity can be inferred by measuring the phase difference
between the beams at two different wavelengths, using the knowledge of the refractive index of wet
air across the infrared. In practice, a blind phase correction in the science band, relying for
instance on H- and K-band measurements, is dangerous because it is not sensitive to possible
biases, e.g.\ related to the imperfect knowledge of the refraction index or to the presence of
other dispersive gases~\citep{Meisner02}. Phase measurements in the science band are thus necessary
to ensure a good co-phasing, using a part of the scientifically useful signal. These measurements
will be carried out at a slower rate, on one hand because of the lower signal-to-noise ratio on
phase measurements at longer wavelengths and on the other hand because the possible biases related
to the extrapolation of the phase delay from the H/K bands towards longer wavelengths are mainly
affected by global atmospheric parameters (temperature, partial pressures of gases, etc.) which are
not expected to evolve quickly during the night. Another possibility would be to perform both
fringe tracking and dispersion control by using science-band photons exclusively, in order to
reduce the technical complexity of the instrument. This would however lead to slightly reduced
performance, due to the worse signal-to-noise ratios on phase measurement at longer wavelengths,
and would even become impractical in the N band.

Interferometric facilities are currently not equipped with a dispersion compensation device, which
should thus be integrated in the nuller itself. The required correction is two-fold: the inter-band
phase error should be corrected with a dedicated short-stroke delay line (less than one $\mu$m RMS
to be compensated), while intra-band dispersion could be corrected by introducing a variable amount
of dispersive material in the beams to correct for the slope of the phase inside the science
waveband~\citep{Koresko02}. The performance of such a correction system is summarized in
Table~\ref{tab:nullperf} in the L' band, once again in two cases:
\begin{itemize}
\item the {\em worst case} is obtained by assuming that phase measurements in the L' band are
needed at each step of the control process, and assuming a $-8/3$ slope for the high-frequency
asymptote of the atmospheric turbulence PSD.
\item the {\em best case} is computed assuming that H/K band phase measurements are sufficient for
dispersion control and that the input PSD has a $-17/3$~slope at high frequencies.
\end{itemize}
The closed-loop performance ranges between 4.4~nm and 17~nm RMS for inter-band dispersion, between
1.0~nm and 4.1~nm RMS for intra-band dispersion. It is marginally compliant with the requirement
(4~nm RMS).

    \subsection{Intensity matching}

The third real-time control loop to be implemented in the nulling instrument is a device to correct
for unequal intensities between the two beams. In fact, intensity fluctuations are not a problem
per se: it is only the differential fluctuations between the beams that matter. The most
straightforward way to compensate for this effect is to measure the intensities of the two beams,
determine which one is the brightest and reduce its intensity down to the level of the other one by
means of an actuator. This correction must take into account the effect of coupling into
single-mode fibers, so that the intensity sensors must also be placed after the modal filters.
Furthermore, it must be carried out in the science waveband, because coupling efficiency is
wavelength-dependent. We will thus ``waste'' an additional part of the useful signal for intensity
control. A possible actuator could consist in a piezo-driven adjustable iris, which would reduce
the beam size in a pupil plane while keeping it (almost) circular to preserve injection properties
as much as possible.

The closed-loop control performance is summarized in Table~\ref{tab:nullperf} for the L' band, with
a typical residual intensity mismatch of 4\% when the loop is running at 1~kHz. This performance
does not meet the required 1\% mismatch, which makes additional calibrations mandatory (see
Sect.~\ref{sub:instcal}). Note that this control loop does not remove the fluctuations of the mean
coupling efficiency (common to both beams), which has a typical standard deviation of 8\%.

    \subsection{Instrumental nulling performance} \label{sub:perf}

\begin{figure}[!h]
\centering \resizebox{\hsize}{!}{\includegraphics{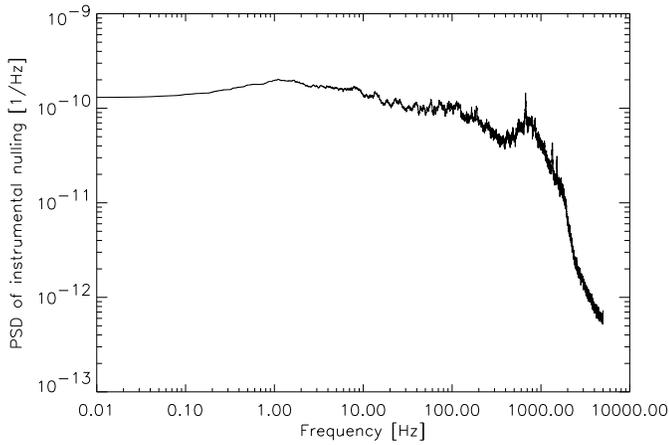}} \caption{Power spectrum of
instrumental nulling due to phase and intensity errors after closed-loop control, obtained for a
100~sec simulation with GENIEsim in the L' band. The semi-analytical simulation takes into account
the bilinear error terms discussed by~\citet{Lay04}, but not the possible $1/f$-type noises
discussed by~\citet{CB05}. The power spectrum has been smoothed with a standard IDL routine to show
its mean behaviour: almost flat at low frequencies, while the logarithmic slope at high frequencies
is about $-2$.} \label{fig:nullpsd}
\end{figure}

Table~\ref{tab:nullperf} shows the nulling performance associated with the residual phase and
intensity errors after closed-loop control for a G2V star at 20~pc. The RMS instrumental nulling
ratio corresponds to the concept of variability noise~\citep{CB05}. Its power spectrum is
illustrated in Fig.~\ref{fig:nullpsd}, showing an almost flat spectrum for frequencies below 1~kHz
and a rapid decrease of the power content above 1~kHz. Variability noise thus behaves as a white
noise at typical read-out frequencies ($\sim$10~Hz), so that its cumulated contribution increases
as the square root of integration time just as for shot noise. We do not take into account here the
possible $1/f$-type noises~\citep{CB05}, which might alter the slope of the PSD at low frequencies
and thus change the behaviour of long-term variability noise by introducing more power at low
frequencies. This assumption is supported by the use of real-time control loops, which will help
remove possible drifts in the instrumental response.

\begin{table}[!h]
\begin{center}
\begin{tabular}{cccc}
\hline \hline & worst case & best case & goal
\\ \hline Piston & 17nm @ 20kHz & 6.2nm @ 13kHz & $<4$nm
\\ Inter-band & 17nm @ 200Hz & 4.4nm @ 300Hz & $<4$nm
\\ Intra-band & 4.1nm @ 200Hz & 1.0nm @ 300Hz & $<4$nm
\\ Intensity & 4\% @ 1kHz & 4\% @ 1kHz & $<1$\%
\\ \hline Total null & $9.7 \times 10^{-4}$  & $6.2 \times 10^{-4}$ & $4.8 \times 10^{-4}$
\\ Instr. null & $5.0 \times 10^{-4}$ & $1.5 \times 10^{-4}$ & $10^{-5}$
\\ RMS null & $5.8 \times 10^{-4}$ & $1.6 \times 10^{-4}$ & $10^{-5}$
\\ \hline
\end{tabular}
\caption{Performance of an L' band nuller at VLTI as simulated with GENIEsim on a 100~sec
observation block, taking into account all turbulence-induced errors. The performance of the
control loops are summarized together with their repetition frequencies in pessimistic and
optimistic cases (see text). The target performance discussed in Sect.~\ref{sec:turb} appears in
the last column. The total null is the mean nulling ratio including both the geometric and
instrumental leakage contributions, and is computed as the ratio between the total input and output
stellar fluxes (from both telescopes). The last line gives the standard deviation of the
instrumental nulling ratio for this 100~sec observation block.} \label{tab:nullperf}
\end{center}
\end{table}

\section{Post-processing of nulling data} \label{sec:calib}

Because the expected performance of state-of-the-art real-time control loops is not sufficient to
reduce instrumental leakage down to the required level ($10^{-5}$ in the L' band), and because
other sources such as geometric leakage and background also contribute to the detected signal after
destructive beam combination, the exozodiacal disk signal can only be extracted from the raw data
after a final and critical phase of correction: post-processing.

    \subsection{Background subtraction} \label{sub:chopping}

The first step in the data analysis procedure is to remove the contribution of the background. In
order to extract the useful data from the incoherent background in the nulled data, modulation
techniques are required. The classical ``chopping-nodding'' technique consists in tilting the
secondary mirror by a small quantity (a few arcseconds) in order to measure the background in an
empty sky region close to the source and subtract it from the on-source measurement. The chopping
sequence is repeated at a few Hertz. Nodding consists in repeating the whole chopping sequence
off-source, by slightly tilting the telescope, in order to remove the possible gradient in the
background emission.

Because of the large temporal fluctuations of the background emission, chopping at a few Hz is not
sufficient to reduce sky noise below the level of the scientifically useful signal after background
subtraction. Other techniques are therefore considered in order to measure the background emission
simultaneously to the observations, e.g.\ by inserting additional fibers in the transmitted
field-of-view of the interferometer. At VLTI for instance, the use of variable curvature mirrors in
the delay lines provides a clean field-of-view 2'' in diameter, which allows us to reduce the
contribution of stellar light in the background fibers down to about $10^{-4}$ of the background
contribution. Another solution for background subtraction is to use phase modulation techniques,
which require the telescope pupils to be divided into two parts in order to modulate between two
different nulled outputs~\citep{Serabyn04}. This is the only practical solution for an N-band
nuller, for which the contrast between background and exozodiacal fluxes is huge (almost $10^7$,
see Table~\ref{tab:bckg}).

    \subsection{Geometric leakage calibration} \label{sub:geomcal}

An important step in the post-processing of nulling interferometry data is the calibration of
geometric leakage. We have discussed in Sect.~\ref{sub:criterion} that the residual contribution of
geometric leakage after calibration should not exceed $10^{-5}$ of the initial stellar flux in the
L' band, while typical rejection rates of 2000 are expected (Table~\ref{tab:rejrate}). The
calibration of geometric leakage, based on the analytical expression~(\ref{eq:rejrate}) of the
rejection rate, should therefore reach a precision of about 2\% in order to retrieve the useful
exozodiacal disk signal with a good signal-to-noise ratio. Closer stars, for which the rejection
rate is smaller, require even better calibration accuracies. {\em A typical precision of 1\% for
geometric leakage calibration is thus required in the L' band}, while such a calibration is
marginally needed in the N band since the achievable rejection rate ($\sim$16000) already matches
the requirement on starlight rejection ($10^{-4}$).

Assuming that the baseline $B$ and the effective wavelength $\lambda$ are known with a good
accuracy, a good knowledge of the stellar angular radius $\theta_{\ast}$ and limb-darkening
coefficient $u_{\lambda}$ is sufficient in order to compute the geometric leakage. The global
optical transmission of the interferometer must also be known in order to convert this value into a
number of photo-electrons measured at the detector. While the interferometric transmission can be
straightforwardly measured with a high precision at the constructive output of the nuller where
most of the stellar photons end up, the imperfect knowledge of stellar diameters has a large (and
generally dominant) contribution to the calibration error budget. Differentiating
equation~(\ref{eq:rejrate}), we obtain a required precision $\Delta \theta_{\ast} / \theta_{\ast} =
0.5 \Delta N / N = 0.5\%$ on the stellar diameter knowledge.

The most precise way to determine stellar angular radii is currently based on stellar
interferometry. Stellar diameter measurements with an accuracy as good as 0.2\% have been
demonstrated on resolved stars such as $\alpha$~Cen A and B \citep{KT03}. However, the Darwin/TPF-I
targets, with typical angular diameters of 1~mas and below, are only marginally resolved even with
the longest baselines currently available: the AMBER instrument at VLTI provides at best an angular
resolution of 1.6~mas in J band. A precision of a few 0.1\% on diameter measurements with AMBER
could thus only be achieved for the closest stars in the Darwin catalogue ($< 10$~pc). An
additional difficulty comes from the extrapolation of the limb-darkening law from the J band to the
L' band, since limb-darkening coefficients are currently modelled only up to the K
band~\citep{Claret00}. Lunar occultation measurements could also be used to determine the diameters
of the brightest targets of the Darwin catalogue ($K\sim 3$) with a limiting resolution of $\sim
1$~mas, but with the restriction that only about 10\% of the sky falls on the Moon's path
\citep{Fors04}.

For stars more distant than about 10~pc, for which interferometric diameter measurements will
probably not reach a 0.5\% precision, indirect measurements based on surface brightness relations
might be used. Empirical laws make it possible to predict the limb-darkened angular diameters of
dwarfs and subgiants using their dereddened Johnson magnitudes, or their effective temperature
\citep{KT04}. Intrinsic dispersions as small as 0.3\% are obtained for the relations based on B$-$L
magnitudes. However, the apparent magnitudes of the Darwin/TPF-I targets need to be accurately
known (to better than 0.02~mag): this is usually not the case at infrared colours and might mandate
a dedicated photometry programme.

All in all, it is expected that diameter measurements with a precision of 0.5\% will be obtained on
most of the Darwin/TPF-I targets. A 1\% precision is still considered as a worst case scenario in
the performance budget.

    \subsection{Instrumental leakage calibration} \label{sub:instcal}

After calibration of geometric leakage, there is one remaining step of calibration to be carried
out: the evaluation and subtraction of the mean instrumental leakage from the nulled data.
Calibration of the instrumental response is a routine procedure in stellar interferometry, and is
achieved by observing a well-know {\em calibrator star}, usually chosen as a late-type unresolved
giant star \citep{Merand05}. These measurements are carried out just before and after the
observation of the target star, and as close as possible to it in the sky, so that the atmospheric
properties do not change significantly.

This method can be extended to measure instrumental leakage in nulling interferometry. In this
case, calibrator stars must be carefully chosen: even if instrumental leakage does not depend on
the angular diameter of the star to the first order, the control loop performance strongly depends
on the stellar flux and spectrum, so that the calibrator star should in fact have the same flux and
spectrum as the target star in order to get the same instrumental leakage in both cases. Having the
same flux and spectrum means that the angular diameters should also be the same, and consequently
that the calibrator star will produce the same amount of geometric leakage as the target star. A
precise knowledge of the calibrator's angular diameter is thus needed in order to remove the
contribution of geometric leakage and obtain a good estimation of instrumental leakage.

The final estimation $Z(\lambda)$ of the exozodiacal signal after both geometric and instrumental
leakage calibrations can be expressed as follows:
\begin{equation}
Z(\lambda) = S_{\rm t}(\lambda) -  \frac{\pi^2 B^2 \theta_{\rm t}^2}{4\lambda^2} F_{\rm t}(\lambda)
- \left( S_{\rm c}(\lambda) -  \frac{\pi^2 B^2 \theta_{\rm c}^2}{4\lambda^2} F_{\rm c}(\lambda)
\right) \; ,
\end{equation}
with $S_{\rm t}(\lambda)$ and $S_{\rm c}(\lambda)$ the background-subtracted nulled fluxes for the
target and calibrator stars, $\theta_t$ and $\theta_c$ their angular radii, $F_{\rm t}(\lambda)$
and $F_{\rm c}(\lambda)$ their fluxes, measured at the constructive output of the interferometer.
The first subtracted term represents the contribution of geometric leakage while the second one
(between brackets) is the estimation of instrumental leakage obtained with the calibrator star.


\section{Detection of exozodiacal dust disks with GENIE} \label{sec:genie}

In the following paragraphs, we evaluate the sensitivity of GENIE, the Ground-based European
Nulling Interferometer Experiment~\citep{Gondoin04}, to exozodiacal dust for typical Darwin/TPF-I
targets.

    \subsection{Background-limited signal-to-noise ratio} \label{sub:bckglimited}

Before computing the actual detection performance of the nuller, let us assess the required
integration times in a non-turbulent environment, where the only source of noise is the shot noise
from all sources inside the field-of-view. The photon budget of Table~\ref{tab:bckg} includes the
geometric and instrumental leakage contributions given in Table~\ref{tab:nullperf} (worst case
scenario), the contribution of the dust disk itself\footnote{The part of the exozodiacal emission
transmitted by the interference fringe pattern shown in Fig.~\ref{fig:tmap} typically ranges
between 30\% and 40\%.} and the different sources of background emission (sky and instrument).

We assume an overall instrumental throughput (including VLTI, GENIE and detectors) of 1\% both in
the L' and N bands. Table~\ref{tab:bckg} gives the time required to detect a 20-zodi disk around a
G2V star at 20~pc with a signal-to-noise ratio of 5. While the integration time in the L' band is
reasonable (15 minutes), it is very large in the N band (11 hours). In fact, the difficulty in the
N band comes not only from the required integration time, but also from the huge contrast between
the exozodiacal and background fluxes (almost $10^7$) at the output of the nuller. These two
background-related issues are the main reasons for the choice of the L' band in the context of
GENIE \citep{Gondoin04}. Note that, even in the L' band, the required background calibration
accuracy is challenging: in order to reach a final signal-to-noise ratio larger than 5, the
background must be removed with an accuracy of $10^{-5}$, which is within reach using the advanced
background subtraction methods discussed in Sect.~\ref{sub:chopping}.

\begin{table}[!h]
\begin{center}
\begin{tabular}{ccc}
\hline \hline & L' & N
\\ \hline Total stellar signal [Jy] & 3.1 & 0.51
\\ Total disk signal [Jy] & $3.0 \times 10^{-4}$ & $5.1 \times 10^{-4}$
\\ Sky brightness [Jy/as$^2$] & 5.0 & 690
\\ VLTI brightness [Jy/as$^2$] & 115 & 21300
\\ GENIE brightness [Jy/as$^2$] & 29 & 5320
\\ Total bckg signal [Jy] & 1.8 & 2550
\\ \hline Final stellar leakage [el/s] & $1.0 \times 10^{4}$ & $1.5 \times 10^{3}$
\\ Final disk signal [el/s] & $3.6 \times 10^{2}$ & $4.1 \times 10^{3}$
\\ Final bckg signal [el/s] & $4.8 \times 10^{6}$ & $2.7 \times 10^{10}$
\\ \hline Shot Noise [el/s$^{1/2}$] & $2.2 \times 10^{3}$ & $1.7 \times 10^{5}$
\\ Time for SNR = 5 [sec] & 950 & 40000
\\ \hline
\end{tabular}
\caption{In the first part of the table, we compare the star and exozodiacal disk fluxes to the
thermal background, computed from its individual contributors (Paranal night sky, VLTI optical
train and GENIE instrument) by integration over the field-of-view defined by a single-mode fiber
($\Omega=\lambda^2/S$). The target is a Sun-like star at 20~pc surrounded by a 20-zodi disk. The
values are given in Jy at the central wavelength of the infrared atmospheric windows and ``at the
entrance of the Earth's atmosphere'' (an equivalent emission is given for the background
contributors as if they originated from outside the Earth atmosphere). Second part: signals in
electrons per second actually detected at the GENIE science detector. The total shot noise is
deduced, and used to compute the time needed for a photometric signal-to-noise ratio of 5.}
\label{tab:bckg}
\end{center}
\end{table}

    \subsection{Final signal-to-noise ratio}

In this section, we evaluate the global sensitivity of the GENIE instrument by simulating a
30-minute observation of a Sun-like star at 20~pc surrounded by a 20-zodi disk. In
Table~\ref{tab:sensitivity}, we first give the raw sensitivity of the nuller, i.e., without
calibration of instrumental leakage (very conservative approach assuming that no convenient
calibrator star can be found). In this case, the mean instrumental leakage fully contributes as a
bias which adds to the exozodiacal disk signal that we want to detect. This implies that the mean
instrumental leakage should be at least five times smaller than the disk contribution to ensure a
safe detection. Table~\ref{tab:sensitivity} shows that, depending on the model for atmospheric
turbulence, the raw sensitivity ranges between 530 and 1800 zodis for a signal-to-noise ratio of 5
across the whole L' band. Instrumental leakage is the main contributor to the noise budget in that
case. Note that shot noise, detector noise and variability noise are almost negligible after only
30~min of integration. This would not be the case if spectral dispersion was used: in order to keep
the same sensitivity, the integration time should then be increased by a factor equal to the square
root of the number of spectral channels.

In a second step, we compute the performance assuming that an appropriate calibrator star can be
found, i.e., with similar flux, spectrum and angular diameter than the target star but without
circumstellar dust. In this case, the calibration precision for instrumental leakage is mainly
limited by the imperfect knowledge of the calibrator angular diameter, assumed to be the same than
for the target star. This contribution is dominant in the noise budget, so that the GENIE
sensitivity could benefit from improvements in the precision of stellar diameter measurements. In
practice, the calibrator star will not be perfectly identical to the target, inducing a bias in the
calibration process due to the different behaviour of the control loops for different stellar
magnitude and/or spectra. This bias is not dominant in the calibration process as long as the H, K
and L' magnitudes of the target and calibrator do not differ by more than about 0.2 magnitudes. A
convenient calibrator will probably not be available in all cases, so that the actual performance
of the nuller could be somewhat degraded with respect to the final values of
Table~\ref{tab:sensitivity} (Absil et al., in preparation).

\begin{table}[!h]
\begin{center}
\begin{tabular}{ccc}
\hline \hline & worst case & best case
\\ \hline 20-zodi signal [e-] & $6.5 \times 10^5$ & $8.7 \times 10^5$
\\ Shot noise [e-] & $1.3 \times 10^5$ & $1.5 \times 10^5$
\\ Detector noise [e-] & $3.4 \times 10^3$ & $4.0 \times 10^3$
\\ Variablility noise [e-] & $1.9 \times 10^4$ & $5.9 \times 10^3$
\\ Calibrated geom. leakage [e-] & $2.2 \times 10^5$ & $1.5 \times 10^5$
\\ Raw instr. leakage [e-] & $1.2 \times 10^7$ & $4.6 \times 10^6$
\\ Calibrated instr. leakage [e-] & $3.8 \times 10^5$ & $2.1 \times 10^5$
\\ \hline Zodis for SNR=5 (raw) & 1800 & 530
\\ Zodis for SNR=5 (calibrated) & 56 & 34
\\ \hline
\end{tabular}
\caption{Expected sensitivity of the GENIE instrument at VLTI, given in number of zodis that can be
detected around a Sun-like at 20~pc in 30~min in the full L' band using the UT2-UT3 baseline
(47~m). The worst case corresponds to conservative control loop performance and 1\% precision on
diameter measurements, while the best case relies on optimistic control loop performance and 0.5\%
precision on angular diameters. Each individual contribution is given in photo-electrons detected
at the nulled output. The raw sensitivity is given without any calibration of instrumental leakage,
while the calibrated sensitivity assumes that a calibrator star with similar characteristics as the
target star is used to perform instrumental leakage calibration.} \label{tab:sensitivity}
\end{center}
\end{table}

    \subsection{Influence of stellar type and distance}

\begin{figure}[!h]
\centering \resizebox{\hsize}{!}{\includegraphics{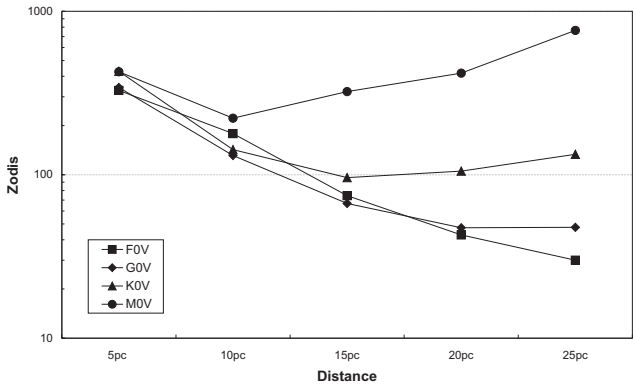}} \caption{Sensitivity of GENIE to
dust disks around nearby main-sequence stars, expressed in units relative to the Solar zodiacal
cloud. The simulations have been obtained with optimized UT-UT baselines for 30~minute exposures.
The smallest baseline (47~m on UT2-UT3) is used in most cases because the dominant noise source
generally comes from the calibration of geometric leakage for both the target and calibrator
stars.} \label{fig:limitUT}
\end{figure}

\begin{figure}[!h]
\centering \resizebox{\hsize}{!}{\includegraphics{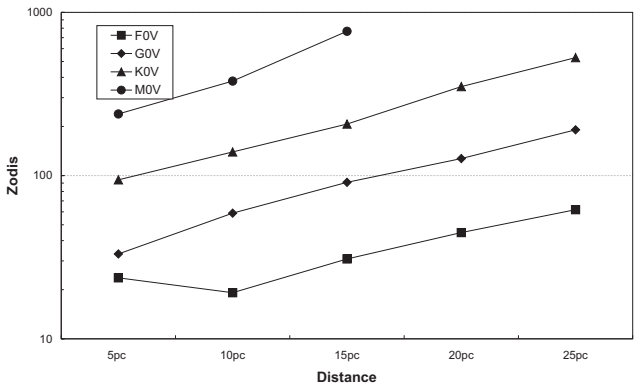}} \caption{Sensitivity of GENIE to
dust disks around nearby main-sequence stars, expressed in units relative to the Solar zodiacal
cloud. The simulations have been obtained with optimized AT-AT baselines in 2~hours of integration.
The availability of very small baselines (down to 8~m) allows us to reduce the contribution of
geometric leakage and thus improves the detection levels of Fig.~\ref{fig:limitUT} at short
distances.} \label{fig:limitAT}
\end{figure}

In order to assess the scientific pertinence of GENIE in the context of the Darwin/TPF-I
programmes, we have evaluated the detection performance for exozodiacal dust clouds around typical
Darwin/TPF-I targets, i.e.\ late-type dwarfs with distances ranging between 5 and 25~pc. In
Fig.~\ref{fig:limitUT} are plotted the expected detection levels for the GENIE instrument with
30~minutes of integration on the 8-m UTs, while Fig.~\ref{fig:limitAT} gives the performance on the
1.8-m ATs for an integration time of 2~hours. The baseline lengths have been optimized in order to
reach an optimum sensitivity to exozodiacal dust, within the limitations imposed by the VLT
interferometer (UT-UT baselines comprised between 47 and 130~m, AT-AT baselines ranging from 8 to
202~m). Here again, we have assumed that a calibrator star similar to the target star can be found
for the evaluation of instrumental leakage. This optimistic assumption is at least partially
compensated by the fact that we use conservative assumptions on atmospheric turbulence properties
and on stellar diameter knowledge (cf.\ the ``worst case'' of Table~\ref{tab:sensitivity}).

The performance estimates in the case of UTs show a maximum efficiency for exozodiacal dust
detection at 25~pc for F0V, 20~pc for G0V, 15~pc for K0V and 10~pc for M0V stars. This maximum is
reached when the two main sources of noise are well balanced: for nearby targets, which are
partially resolved, the contribution of geometric leakage is dominant, while for more distant and
thus fainter targets the main contribution comes from shot noise. The strong limitation on the
sensitivity for close targets can be overcome by using smaller baselines, which are available only
for AT pairs at the VLTI. This is illustrated in Fig.~\ref{fig:limitAT}: with the shortest
baselines, one can achieve exozodiacal dust detection down to the 20-zodi level for bright and
close targets (especially F-type stars), which are however scarce in the Darwin star catalogue.

These simulations indicate that exozodi detection down to a density level of about 50 times our
local zodiacal cloud is a realistic goal for most F and G stars in the Darwin catalogue, while
exozodiacal densities of about 100~zodis could be detected around K-type stars. In the case of M
stars, only the closer stars could lead to a pertinent detection level, of about 200 zodis.

\section{Conclusions and perspectives}

In this paper, we have demonstrated the exceptional capabilities of ground-based nulling
interferometers to detect faint circumstellar features such as exozodiacal disks. Ground-based
nulling interferometry requires high-performance closed-loop control of atmospheric turbulence in
the L' band, and extremely precise background subtraction in the N band. In this paper, we have
computed the expected performance of state-of-the-art servo loops for piston, dispersion and
intensity control, allowing the stabilization of the instrumental nulling ratio. Calibration
procedures have been discussed to further remove stellar light from the nulled output of the
instrument. This comprehensive study shows that an L'-band nuller such as GENIE at VLTI could
detect exozodiacal disks about 50 to 100 times as dense as the Solar zodiacal cloud. This
instrument will thus significantly improve the detection performance of current infrared and
sub-millimetric facilities, and above all will peer into the currently inaccessible warm region of
dust disks located within a few AU of nearby late-type dwarfs (typical Darwin/TPF-I targets).

The simulated performance of the nulling instrument critically depends on a number of noise
contributions which could be improved in different ways. Geometrical leakage could be reduced by
using short baselines, typically ranging between 8 and 40~m, while its calibration would be
improved by obtaining accurate stellar diameter measurements at long baselines ($>300$~m) and/or
precise photometry to derive diameters from surface-brightness relationships, and by extending the
limb-darkening models towards the mid-IR wavelengths used by the nulling instruments. Instrumental
leakage depends strongly on the ability to correct the effects of atmospheric turbulence, and would
therefore benefit from advances in real-time control algorithms and compensation devices.
Polarization errors, which have been discussed only briefly in this paper, could also be critical
and need to be carefully addressed at the design level. Finally, advanced background subtraction
methods such as the phase modulations techniques foreseen for the Darwin/TPF-I missions might
further improve the overall performance of a ground-based nuller.

\begin{acknowledgements}
The authors wish to thank their ESA and ESO colleagues, the two industrial partners (Alcatel Space
and EADS-Astrium) and the GENIE scientific team for their support and contributions. The authors
are also grateful to the anonymous referee and to Jean Surdej for their precious help in improving
the quality and readability of the paper.
\end{acknowledgements}


\bibliographystyle{aa} 
\bibliography{3516} 

\end{document}